%% file: main.tex
\documentclass[12pt]{article}

\include{_preamble}
\usepackage[colorlinks,citecolor=blue,urlcolor=blue]{hyperref}

\title{\titlepaper}

\usepackage[normalem]{ulem}
\begin{document}
\maketitle

\begin{abstract}
There is a long-standing debate in the statistical, epidemiological and econometric fields as to whether nonparametric estimation that uses data-adaptive methods, like machine learning algorithms in model fitting, confer any meaningful advantage over simpler, parametric approaches in real-world, finite sample estimation of causal effects. We address the question: when trying to estimate the effect of a treatment on an outcome, across a universe of reasonable data distributions, how much does the choice of nonparametric vs.~parametric estimation matter? Instead of answering this question with simulations that reflect a few chosen data scenarios, we propose a novel approach evaluating performance across thousands of data-generating mechanisms drawn from non-parametric models with semi-informative priors. We call this approach a Universal Monte-Carlo Simulation. We compare performance of estimating the average treatment effect across two parametric estimators (a g-computation estimator that uses a parametric outcome model and an inverse probability of treatment weighted estimator) and two nonparametric estimators (Bayesian additive regression trees and a targeted minimum loss-based estimator that uses an ensemble of machine learning algorithms in model fitting). 
We summarize estimator performance in terms of bias, confidence interval coverage, and mean squared error. We find that the nonparametric estimators nearly always outperform the parametric estimators with the exception of having similar performance in terms of bias and similar-to-slightly-worse performance in terms of coverage under the smallest sample size of N=100. 
\end{abstract}

\noindent%
\vfill

\section{Introduction}
he past two decades have seen rapid growth of nonparametric
statistical estimation methods \cite[e.g.,][]{van2006targeted,chernozhukov2018double,hill2011bayesian,lecun2015deep,hahn1998role}. A large and growing
subset of the statistical,  epidemiological,
econometric, and other applied literatures take as a given that
nonparametric estimation methods are superior to parametric
approaches, particularly in terms of reducing or eliminating model misspecification bias, which can be substantial in an over-simplified parametric model. (Minimizing all sources of bias is essential, especially because when bias remains on the same order as the standard error, the probability of a corresponding hypothesis test rejecting the null will tend to one even when no effect is present.) However, few have contributed evidence of the extent to
which this assumption of nonparametric estimator superiority is borne out in real-world, finite-sample
analyses, leading to debate as to whether nonparametric methods confer
any meaningful practical advantage in typical applied estimation of
causal effects \cite{imbens2004nonparametric}.

Both nonparametric and parametric estimation methods generally involve using theory and
subject-matter knowledge to inform the underlying causal model/graph \cite{pearl2009causality} and
the variables input into the model \cite{zhao2019causal}. However,
whereas parametric estimation methods would proceed by specifying
parametric models (e.g., linear regression) for each component of the causal model used in estimation (e.g., the outcome model), specifying its functional form, interactions, higher-order variable forms, etc., nonparametric methods would typically involve using data-adaptive
methods, like machine learning algorithms in flexibly fitting model
components. Because correctly specifying parametric models \textit{a-priori} seems unlikely in the absence of a clear mechanistic understanding, nonparametric methods can guard against bias
due to model misspecification.

Consequently, nonparametric methods may be expected to have an
advantage when an accurate model is important to obtain an unbiased
estimate, e.g., in the presence of significant, complex confounding or
censoring, especially when response and/or treatment assignment surfaces are
nonlinear \cite{dorie2019automated}. In real data analyses, this may happen with observational study data with strong, complex confounding of the exposure-outcome relationship, or in both observational study data and randomized control trial data with significant, differential dropout, mediation questions, or estimation of heterogeneous treatment effects \cite{wendling2018comparing}.

Previous simulation studies comparing performance between parametric and
nonparametric estimation have generally shown that nonparametric
estimators incorporating data-adaptive, machine learning algorithms
outperform parametric approaches, especially under the data structures
enumerated
above \cite{dorie2019automated,porter2011relative,ozery2018adversarial}. However, common critiques of simulation
studies are that they (a) are typically designed to illustrate a difference in performance (or lack thereof), and thereby reflect hand-picked settings that are favorable to the method of interest; (b) are oversimplified ``stylized models of reality'';
and (c) may hold little value if the goal is to make a general statement
about the degree to which the choice of nonparametric versus
parametric estimation matters \cite{parikh2022validating,schuler2017synth,advani2019mostly,huber2013performance,busso2014new}. A similar criticism regarding lack of
generalizability can apply to the few previous real data analysis
examples comparing nonparametric versus parametric estimators, sometimes finding meaningful differences and other times, not \cite{kreif2019machine,keele2018comparing}.

Responding to the critique that simulation-based evaluations of estimator performance have been conducted on oversimplified data too far removed from the complexity of real-world datasets, two groups recently proposed more comprehensive, yet tailored, evaluations of estimator performance by simulating datasets that closely mimic an observed dataset in all its complexity (similar to the idea of ``plasmode simulation'' \cite{franklin2014plasmode}) \cite{schuler2017synth,parikh2022validating}. This work is premised on the belief that the optimal estimator will differ for different data scenarios, hence the focus on generating simulated data to match a particular observed dataset. However, by using the same data to both select an estimator (on the basis of its performance on a new data-generating mechanism based on the observed data), and evaluate this estimator, these approaches are vulnerable to problems with post-selection inference. Given sufficiently large sample sizes, this problem can be circumvented via sample splitting; however, practitioners can seldom afford to discard large amounts of data to perform estimator selection.


The current debate about how much the choice of
nonparametric versus parametric may matter in the real world asks a
more general question than the current literature can
answer. Namely, across a universe of reasonable data distributions, how
much does the choice matter? We propose an approach to answer such a
question: a Universal Monte-Carlo Simulation that instead of considering just a few data-generating mechanisms, summarizes estimator performance across thousands of data-generating mechanisms. Unlike the approaches of \cite{schuler2017synth,parikh2022validating}, ours does not depend on the observed data. 
We apply this approach to provide a general
quantification of the degree to which choosing a nonparametric versus
parametric estimation approach impacts bias, confidence interval coverage, and mean squared error in the resulting causal effect estimates.


This paper is organized as follows. In Section 1, we introduce notation, the causal estimand we consider, and summarize the necessary theory underlying parametric and nonparametric approaches for estimating the causal estimand. 
In Section 2, we describe our proposed Universal Monte-Carlo Simulation for comparing parametric and nonparametric
estimation. Also in this section, we provide the specific parameters that define the universe of data-generating mechanisms we consider, the parametric and nonparametric estimators we compare, and how we quantify and summarize estimator performance in finite samples. In Section 3, we provide and discuss results. 
Section 4 concludes.



\section{Notation, estimands, and background on parametric and nonparametric estimation}
\label{sec:theory}

We focus on
estimation of the average treatment effect (ATE) in our simulations. The ATE is the average difference in the expected outcomes under treatment versus under control. For simplicity, in this section, we focus on one component of the contrast---
the expected counterfactual outcome had treatment been set to some value, $t$, possibly contrary to fact, denoted $\E(Y^{t})$, where $T$ denotes a binary treatment variable, and $Y$ denotes the outcome variable. The ATE would then be denoted $\E(Y^{1}) - \E(Y^{0})$. We assume that
the exchangeability assumption holds, $Y^{(t)}\indep T\mid X$ for $t\in\{0,1\}$, where $X$ denotes confounding variables. This is
necessary for the causal parameter, $\E(Y^{(t)})$, to be identified from the observed data, $Z=(X,T,Y)$, by the parameter $\theta(\P) = \E[\E(Y\mid T=t,X=x)]$, where $\theta$ denotes the parameter, which is a function of the data distribution, $\P$. 

We note that $\theta(\P)$ constitutes the building block for estimation of common marginal causal effects: the ATE, identified by $\E\{\m(1,X) - \m(0,X)\}$); the relative risk, identified by $\frac{\E\{\m(1,X)\}}{\E\{\m(0,X)\}}$; and the odds ratio, identified by $\frac{\E\{\m(1,X)\}/\E\{1-\m(1,X)\}}{\E\{\m(0,X)\}/\E\{1-\m(0,X)\}}$, where $\m(t,x)$ represents $\E(Y\mid T=t,X=x)$. Each of these parameters can
be represented as a functional mapping a probability distribution $\P$
to the real numbers, and can be denoted with $\theta(\P)$. 

When the exposure takes values in a discrete
set, like the binary exposure we consider here, the above parameters can be nonparametrically estimated, yet with the same asymptotic properties as if they were parametrically estimated, as
long as the dimension of $X$ is fixed. 
Taking the $\ate$ as an example, this means that it will be
possible to find an estimator $\widehat\ate$ such that as the sample
size $n$ grows, $\sqrt{n}(\widehat\ate - \ate)$ converges to a random
variable that is normally distributed with mean zero and variance
equal to the non-parametric efficiency bound. The efficiency bound is the smallest possible variance attainable by a regular estimator, see \cite{hahn1998role}. This result is useful, because it
means that the normal distribution can be used to approximate the
sampling distribution of the estimator $\widehat\ate$ for finite
sample sizes, which allows us to construct approximately correct
confidence intervals and hypothesis tests. 

\subsection{Parametric estimation with substitution estimators}
\label{sec:para}
We can alternatively write $\theta(\P)$, defined  above, as:
\begin{equation}
  \label{eq:deftheta}
  \theta(\P) = \int \m(t,x)\dd\P(x).
\end{equation}
A natural estimator for this parameter is given by the following procedure:
\begin{enumerate}
\item Fit a model for $\m(t,x)$, e.g., the logistic regression model
  $\logit \m_\beta(t,x) = \beta_0 + \beta_1t + \beta_2^\top x$. Let
  $\hat\beta$ denote the maximum likelihood estimate (MLE).
\item For each subject $i=1,\ldots,n$, compute the predicted outcome
  $\m_{\hat\beta}(t, X_i)$ in a hypothetical world where their treatment level
  is set to $T_i=t$ for everyone. For example, for the logistic
  regression model above this is
  $\m_{\hat\beta}(t,X_i) = \expit(\hat\beta_0 + \hat\beta_1t +
  \hat\beta_2^\top X_i)$, where $\expit(x)=\{1 + \exp(-x)\}^{-1}$.
\item Compute the substitution estimator of $\theta(\P)$ as
  \[\thetasub = \theta(\hat \P) = \frac{1}{n}\sum_{i=1}^n\hat\m_{\hat\beta}(t, X_i).\]
\end{enumerate}
This estimator is often referred to as the g-computation estimator in the biostatistics and epidemiology literatures. Note that we have denoted this estimator with $\theta(\hat\P)$, since
it is the result of plugging in a regression estimate of $\m(t,x)$ and
the empirical distribution estimate of $\P(x)$ in the definition of
the parameter (\ref{eq:deftheta}). 

Even when the parametric logistic regression model is
wrong, it can be used to estimate parameters
such as the $\ate$, $\mrr$, and $\mor$. This is
not only true for logistic regression; any regression fit
$\hat\m(t,x)$ (including machine learning) can be mapped to an
estimator of $\theta(\P)$ using the above procedure. 

When the regression $\m(t,x)$ is estimated with a parametric model
(e.g., logistic, linear, Poison regression), standard asymptotic tools such
as the delta method yield the following result. Let $\bar\beta$ denote
the probability limit of the MLE $\hat\beta$
of the model parameters, and let $\m_\beta(t,x)$ denote the model fit
for any value $\beta$. Then, we have $\thetasub \approx N\left\{\theta(\P) + \Bsub(\P), \sigmasub^2/n\right\}$. 
We let $\Bsub(\P)$ denote the asymptotic bias of the estimator ($\Bsub(\P) = \int(\m_{\bar \beta}(t, x) - \m(t, x))\dd\P(x)$),
and $\sigmasub^2\geq 0$ denote the asymptotic variance. We note that the bias
depends on $\P$ through three quantities: the limit $\bar\beta$ of the
model parameters, the true function $\m(t,x)$, and the probability
distribution of the covariates $\P(x)$. The approach we propose in Section \ref{sec:approach} allows us to summarize the distribution of $\Bsub(\P)$ (which depends on $\P$, but is fixed for a given $\P$) across a wide range of data-generating mechanisms, $\P$.  

\subsection{Nonparametric estimation using machine learning for model fitting}\label{sec:ml}
When the regression $\m(t,x)$ is estimated with a machine learning
(data-adaptive) algorithm, the delta method cannot be applied and
other tools must be used for inference. Specifically, the substitution estimator that uses machine learning in model fitting,
$\theta(\hat\P)$, has the following first-order bias: $-\E[\phi(Z;\hat\P)]$, where $\phi(Z;\P) = \frac{\one\{T=t\}}{\g(t\mid X)}\{Y-\m(t, X)\} + \m(t,X)
  -\theta(\P),$ and $\g(t\mid x)=\P(T=t\mid X=x)$, is the propensity score \cite{rosenbaum1983central}. 
  
This bias expression is at the core of frequentist
nonparametric efficient estimators, 
 such as targeted
minimum loss based estimation
\citep[TMLE][]{van2011targeted,van2018targeted} and double/ debiased
machine learning \citep{chernozhukov2018double}. For example, TML estimators are constructed such that
the above bias is approximately equal to zero. The double/debiased ML
estimators are focused on obtaining a reasonable estimate of the bias
and subtracting it from the substitution estimator.  For example, the
one-step estimator (in this case, also known as the augmented inverse probability
weighted estimator) may be constructed as:
\begin{equation}
  \thetaos = \theta(\hat \P) + \frac{1}{n}\sum_{i=1}^n\phi(Z_i,\hat
  \P),  \label{eq:os}  
\end{equation}
and the TMLE is equal to $\thetatmle = \theta(\tilde \P)$, where
$\tilde \P$ is such that
\begin{equation}
  \frac{1}{n}\sum_{i=1}^n\phi(Z_i,\tilde
  \P)\approx 0.  \label{eq:tmle}
\end{equation}
 Under regularity conditions\footnote{Some of
  these conditions can be avoided by cross-fitting the estimators of
  $\g(t\mid x)$ and $\m(t, x)$. That is, the training data used to
  obtain $\hat\g(t\mid X_i)$ and $\hat\m(t, X_i)$ should not contain
  observation $i$.} it can be shown that if the estimators $\hat\m$
and $\hat\g$ converge to some functions $\bar \m$ and $\bar \g$,\footnote{Specifically, this assumes convergence in $L_2(\P)$-norm at $n^{1/4}$-rate.} then the one-step and TML estimators are doubly robust. That is, $\thetatmle \xrightarrow{p} \theta(\P) + \Btmle(\P)$, where 
$\Btmle(\P)=\int \left(\frac{\g(t\mid x)}{\bar\g(t\mid x)} -
    1\right)\left(\m(t,x) - \bar\m(t,x)\right)\dd\P(x)$
is the asymptotic bias, which is zero if either $\bar{\g}=\g$ or $\bar{\m}=\m$, i.e., if either the outcome regression or propensity score estimator is consistent. When both equalities hold (i.e., both are consistent), these estimators are asymptotically normal and semiparametric efficient, meaning
$\sqrt{n}\{\thetatmle - \theta(\P)\} \rightsquigarrow N\left(0,
  \sigma_{\text{eff}}^2\right),$
where $\sigma_{\text{eff}}^2$ is the semiparametric efficiency bound. The asymptotic variance $\sigma_{\text{eff}}^2$ can be consistently estimated by the sample variance of $\phi(Z_i,\hat{\P})$. 

However, we note that nonparametric Bayesian estimators are not constructed around this bias term, and instead use flexible Bayesian formulations for modeling coupled with a plug-in estimation strategy. 

\subsection{Theoretical comparison of parametric and nonparametic estimator bias}

$\Btmle(\P)$ is an integral of a
product of two regression errors. If the two errors are small, this
product will generally be smaller than the single error involved in
the definition of $\Bsub(\P)$. Thus, it will often be the case that
$|\Btmle(\P)|\leq |\Bsub(\P)|$. Use of machine learning to
fit $\m$ and $\g$ will typically make the regression errors involved in
$\Btmle(\P)$ even smaller. This may also be true for the bias of
$\thetasub$, but substitution estimators based on na\"ive machine learning are
not generally asymptotically normal, which challenges the construction of sampling distributions, and so is problematic for frequentist statistical inference (though we note it is 
not a problem for Bayesian statistical inference). Because $\Btmle(\P)$ involves the product of two regression errors, it will be zero if
either $\m$ or $\g$ is known, a property known as \emph{double
  robustness}.

In randomized trials, $\g(t\mid x)$ is known by design, and therefore
$\Btmle(\P)$ (unlike $\Bsub(\P)$) can be made exactly zero
due to double robustness. 
Even in the randomized trial setting, covariance adjustment estimators based on modeling $\theta(\P)$ are typically preferred to unadjusted estimators based on $\E(Y\mid T=t)$, because covariance adjustment estimators
can use baseline characteristics to control for noise in the outcome, thereby yielding more precise effect
estimates. 

Based on these asymptotic results, nonparametric estimators of $\theta(\P)$ are expected to have better performance than parametric estimators in very large samples. 
 However, asymptotic approximations may be poor in small or even moderately-sized real-world samples that are afflicted by the curse of dimensionality, 
resulting in bias and under-coverage of confidence intervals \citep{robins1997toward}. 
The sample size at which asymptotic approximations reflect reality is generally problem-dependent and unknown. 
In the next sections, we develop a simulation-based approach to evaluate if and when nonparametric estimators of $\theta(\P)$ can be expected to outperform parametric estimators across a range of finite samples under minimal but reasonable assumptions on the nature of the true data generating mechanism.

\section{A Universal Monte-Carlo Simulation approach to systematically evaluate the finite
  sample performance of estimators}
\label{sec:approach}
We propose the following general method to systematically evaluate
and compare the performance of a set of candidate estimators in finite samples across a largely unrestricted space of data distributions. To define the space of distributions, we assume some information about the data generating mechanism (e.g., in terms of the number and type of variables, confounding bias, treatment effect
heterogeneity, and degree of nonlinearity), drawing on
the nonparametric Bayesian literature.  
\begin{enumerate}
\item For each $j\in\{1,\ldots,J\}$, we draw a probability distribution
  $\P_j$ using a minimally-informative prior across the space of distributions (i.e., nonparametric model), $\mathcal M$. 
   The minimally-informative distribution on
$\mathcal M$ reflects the fact that investigators often have little
knowledge of the data generating mechanism before seeing the
data. We describe our procedure in
  \S\ref{sec:dgm}.
\item Then, for each of the $J$ data generating mechanisms, we generate $S$ simulated datasets. So, for each $j,s$ we have a dataset $\mathsf D_{j,s}$ of some finite sample size
  $n$ from $\P_j$. 
\item Then, for each of $L$ estimators being considered, we compute the estimate $\hat{\theta}_{l,j,s}$. Let $\hat{\theta}_{1,j,s},\ldots,\hat{\theta}_{L,j,s}$ denote
  the estimates.
\item We next compute performance metrics based on sample averages across the $S$ simulated draws. For distribution
  $\P_j$ and sample size $n$,  the Monte Carlo bias, confidence interval coverage, and MSE of a given estimator
  $\hat\theta_l$ are
  \begin{align*}
    \Bn(\P_j) &= \frac{1}{S}\sum_{s=1}^S \hat\theta_{l,j,s} -
                    \theta(\P_j)\\
    \Covn(\P_j) &= \frac{1}{S}\sum_{s=1}^S \mathbbm{1}\bigg\{\hat\theta_{l,j,s} - \frac{c \hat{\sigma}_{l,j,s}}{\sqrt{n}} < \theta(\P_j) \\
    &< \hat\theta_{l,j,s}  + \frac{c \hat{\sigma}_{l,j,s}}{\sqrt{n}}\bigg\}
                    \\
    \MSEn(\P_j) &= \frac{1}{S}\sum_{s=1}^S \{\hat\theta_{l,j,s} - \theta(\P_j)\}^2,
  \end{align*}
  where $c$ denotes the 97.5th percentile of the standard normal distribution.
\item Lastly, we summarize performance across the $J$ data distributions with a probability distribution of each performance metric (e.g., the bias). Specifically, the bias for estimator $l$ at sample
  size $n$ is summarized by
  \[\mathbb P(|\Bn| > b) = \frac{1}{J}\sum_{j=1}^J\mathbbm{1}\{|\Bn(\P_j)| > b\}.\]  This function can be seen as an approximation of the so-called survival or reliability function, where the former name is common in the statistics literature and the latter in the engineering literature. We use the name reliability function, because it is closer to the intended interpretation for each estimation procedure. Because we consider the sample space of probability distributions as all possible data generating mechanisms (i.e., phenomena under study), the probability distribution can be heuristically interpreted as probabilities across all possible studies. 
\end{enumerate}

In contrast to standard simulation studies, which
often focus on a few data generating mechanisms, this Univeral Monte-Carlo Simulation
will provide evidence of the behavior of the estimators across many
data generating mechanisms, and consequently, will be informative for
assessing how the estimators perform across the range of problems that
may be encountered in practice.


\subsection{Specifying and sampling from the space of probability distributions}\label{sec:dgm}

Our simulations will be restricted to binary treatments and binary outcomes. Although we do
allow for non-binary confounding variables, we restrict to discrete, ordinal variables. These restrictions ensure computational tractability but generalizations can in principle be devised for any data structure. Our models for $\P$ will be characterized by the following user-given, minimally-informative priors, representing an analyst's prior knowledge of the data
generating distribution.
\begin{itemize}[noitemsep]
\item An integer $u$ representing the number of binary confounding variables.
\item An integer $h$ representing the number of non-binary confounding variables, together with an integer $c$ denoting the
  cardinality of the support of each. Without loss
  of generality, we assume that each is supported
  in the set
  $\mathcal N = \{0, 1/(c-1), 2/(c-1),
  \ldots, 1\}$.
\item A parameter $\eta$ that controls the non-linearity of the effect of the
  non-binary confounding variables in the data
  generating process (additional detail below).
\item A parameter $\rho$ that controls the smoothness of the effect of the
  non-binary confounding variables in the data
  generating process (additional detail below).
\item An interaction order $k$ (additional detail below).
\item A boolean value $\texttt{hte}\in\{\texttt{TRUE}, \texttt{FALSE}\}$ indicating whether
  there is treatment effect heterogeneity, meaning that the effect of the treatment on the outcome varies by the level of one or more confounding variables.
\item A positive number $q$ bounding the treatment
  probabilities as
  \[\max_{t\in\{0,1\}, x\in \mathcal X}\frac{\P(T=t)}{\P(T=t\mid
      X=x)}\leq q,\]
  where $\mathcal X$ is the support of $X$.
  \item A real number $b$ representing the desired \textit{confounding
      bias}: $\E(Y\mid T=1)  - \E(Y\mid T=0) - \E\{\m(1, X)
      - \m(0,X)\}$.
\end{itemize}

Our sampling scheme proceeds sequentially by first sampling $\P(X=x)$,
then $\P(T=1\mid X=x)$, and finally $\P(Y=1\mid T=t, X=x)$. Note that
in the above setup, the vector $X$ takes values on the set
$\mathcal X = \{0,1\}^{u}\times {\mathcal N}^{u}$, which has
cardinality $2^{u} \times c^{h}$. In the first step, we sample the
vector $\{\P(X=x):x\in\mathcal X\}$ using a Dirichlet uniform distribution in
$[0,1]^{2^{u} \times c^{h}}$. Other distributions could be used that would induce more or less correlation among the covariates; see \cite{dunson2009nonparametric} for distributions over the space of multivariate, categorical data.   

\subsection{Sampling treatment probabilities} Let $X=(\Xb, \Xn)$ represent the partitioning of binary and non-binary confounding variables.  Treatment probabilities are
generated using a linear probability model where we consider:
\begin{enumerate}
\item interactions up to $k$-th order for $\Xb$, and
\item interactions of each of the above terms with a non-linear
  transformation of $\Xn$.
\end{enumerate}
Let $L=\{l =(l_1,\ldots, l_u): l_j\in\{0,1\};\sum_j
l_j\leq k\}$. For instance, when $u=3$ and $k=2$, $L=\{(0,0,0),\allowbreak (1,0,0),\allowbreak (0,1,0),\allowbreak (0,0,1),\allowbreak (1,1,0),\allowbreak (1,0,1),\allowbreak (0,1,1)\}$. Note that this set has cardinality $|L|=\sum_{j=0}^k{u \choose
  j}$. 
Mathematically, the model takes the form $\P(T=1\mid
X=x)=\g(x;\alpha)$, where
\[\g(x;\alpha)=\sum_{l\in L}\{\alpha_{0,l} +
  \alpha_{1,l}f_l(\xn)\}\prod_{j=1}^{u} \xbj^{l_j}.\]
For example, when $u=2$ and $k=1$, 
\begin{align*}
    \g(x;\alpha) =&\;  \alpha_{0,1} + \alpha_{1,1}f_1(\xn) + \alpha_{0,2}x_{\text{bin},1}f_2(\xn)\\
    &+ \alpha_{0,3}x_{\text{bin},1} + \alpha_{1,3}x_{\text{bin},1}f_3(\xn).
\end{align*}
  
The parameters
$(\alpha_{0,l},\alpha_{1,l}, f_l)$ are sampled as follows. First, each
$f_l(\xn)$ is sampled from a Gaussian process with linear mean
$\gamma^{\top}\xn$ and covariance function equal to
\[K(\xni,\xnj)=\eta \exp\{-\rho ||\xni - \xnj||^2\},\] where each
coefficient in $\gamma$ is independently drawn from a standard normal
distribution. Then, the vector of coefficients
$\{(\alpha_{0,l}, \alpha_{1,l}):l\in L\}$ is sampled uniformly from a convex
polytope defined by the following linear constraints for all
$x\in\mathcal X$:
\begin{align*}
  1 &\geq \g(x;\alpha) \\
  0 &\leq \g(x;\alpha)\\
  q &\geq \frac{\sum_x\g(x;\alpha)\P(X=x)}{\g(x;\alpha)} \\
  q &\geq \frac{\sum_x\{1-\g(x;\alpha)\}\P(X=x)}{1-\g(x;\alpha)},
\end{align*}
where the first two constraints ensure that $\g(x;\alpha)$ is a
well defined probability. The third and fourth constraints prevent near-positivity violations governed by the
parameter $q$. That is, they ensure the conditional probability of each treatment level given each covariate value is not too small relative to the marginal probability of that treatment level. Note that the third and fourth constraint are indeed
linear as, for example, the third constraint can be rewritten as
$\sum_x\g(x)\P(X=x) - q\times\g(x;\alpha)\leq 0.$
The above sampling is performed using the \texttt{volesti} R package
\citep{volesti}. 

Once a vector $\{(\alpha_{0,l}, \alpha_{1,l}):l\in L\}$ is sampled, it is
checked for whether it can possibly yield a desired confounding bias
$b$, the details of which are given in the SI Appendix. If not, the current draw is rejected. 

\subsection{Sampling the outcome mechanism}

Outcomes are also generated using a linear probability
model: $\P(Y=1\mid T=t, X=x) = \m(t,x;\lambda, \beta)$, where $\m(t,x;\lambda, \beta) $ \[= t\sum_{l\in L}\{\lambda_{0,l} +
  \lambda_{1,l}h_{l}(\xn)\}\tilde x_l + \sum_{l\in L}\{\beta_{0,l} +
  \beta_{1,l}w_{l}(\xn)\}\tilde x_l\]
if there is treatment effect heterogeneity, and 
\[= t\lambda + \{\beta_{0,l} +
  \beta_{1,l}w_{l}(\xn)\}\tilde x_l\] if there is no treatment effect heterogeneity, where
$\tilde x_l = \prod_{j=1}^{u} \xbj^{l_j}$ to simplify notation. The
functions $h_l(\xn)$ and $w_l(\xn)$ are drawn from Gaussian processes as
before. The confounding bias for a given distribution is equal to
\begin{equation}
\begin{aligned}
     C &= \sum_t(2t-1)\sum_x\frac{\P(X=x)}{\P(T=t)} \left\{\P(T=t\mid X=x) -
    \P(T=t)\right\}\\
    &\times\P(Y=1\mid T=t, X=x),\label{eq:conf}
\end{aligned}
\end{equation}
and is
is linear in the coefficients $\{\lambda_{t,l},\beta_{t,l}:t\in\{0,1\}, l\in L\}$. So, for a tolerance $\texttt{tol}$, we can draw the coefficients
from a uniform distribution in the polytope defined by the following linear
constraints:
\begin{align*}
  1 &\geq \m(t, x;\lambda,\beta) \\
  0 &\leq \m(t, x;\lambda,\beta)\\
  b + \texttt{tol} &\geq C(\lambda,\beta) \\
  b - \texttt{tol} &\leq C(\lambda,\beta),
\end{align*}
where $C(\lambda,\beta)$ denotes equation (\ref{eq:conf}) with
$\P(T=1\mid T=t, X=x)$ replaced by $\m(t, x;\lambda,\beta)$.

\subsection{Our application of the Universal Monte-Carlo Simulation approach}
We now use the above described approach to summarize performance of parametric and nonparametric estimators in estimating the average treatment effect, $\ate =\E\{\m(1,X) - \m(0,X)\}$.

We define our universe of data generating distributions, $\mathcal{M}$, as follows. In addition to one binary treatment and one binary outcome, we considered five binary covariates ($u = 5$) and one numerical covariate ($h = 1$) with cardinality 100. We considered interactions between covariates of order $k=\{1, 2, 3\}$.  We considered distributions both with and without treatment effect heterogeneity ($\texttt{hte}\in\{\texttt{TRUE}, \texttt{FALSE}\}$), and limiting the treatment probabilities as being $\ge 0.001$ ($q = 1000$). The parameter, $\eta$, controlling the nonlinearity of the numerical confounder was sampled from a uniform distribution $U(0.1, 10)$; the parameter, $\rho$, that controls the smoothness of this numerical confounder was sampled from a uniform distribution $U(0.1, 10)$ (see Figure S1 in the appendix to visualize the nonlinearity/smoothness); and the parameter, $b$, that controls the confounding bias was sampled from a uniform distribution $U(-0.3,0.3)$. 

Within this universe of data generating distributions, and for each combination of $k$ and indicator of treatment effect heterogeneity, we sampled $J=500$ distinct DGPs, resulting in $3\times2\times500=3,000$ distributions. For each of these 3,000 distributions, we then sampled $S=250$ data sets for each of the sample sizes $N \in \{100, 500, 1000\}$. In total, we simulated 2,250,000 data sets\footnote{Some DGPs failed to generate after 1000 iterations; the final number of data sets was 1,773,000.}. 

For each data set we then estimated the $\ate$ using 
$L=4$ parametric and nonparametric estimators. 1) First, we considered a parametric substitution estimator (also called a g-computation estimator), as described in Section \ref{sec:theory}, based on modeling the outcome using logistic regression with only main effect terms included. This estimator is commonly used in practice and may be less sensitive than others to the curse of dimensionality. 2) The other parametric estimator we considered is an inverse probability of treatment weighting (IPTW) estimator with weights determined using the covariate balancing propensity score (CBPS), which optimizes the balance of covariates across the treatment and control groups \citep{imai2014covariate}. Propensity score estimators are popular in applied research, despite their inefficiencies \cite{robins2007comment}. The CBPS estimator is one of the best-performing propensity-score-based estimators in finite samples and is robust to mild misspecification of the parametric model \citep{imai2014covariate}. 3) Third, we considered a Bayesian nonparametric estimator: Bayesian adaptive regression trees (BART, \cite{hill2011bayesian, chipman2010bart}), and 4) fourth, we considered the nonparametric TMLE \cite{van2006targeted} as described in Section \ref{sec:theory}, because these nonparametric estimators performed well in many finite sample scenarios and data challenges previously \cite{dorie2019automated}. For data-adaptive estimation in TMLE, we use the Super Learner
\citep{vdl2007super} with a library of estimators consisting of main-effects generalized linear models, BART \cite{chipman2010bart}, light gradient-boosting machine \cite{ke2017lightgbm}, and multivariate adaptive regression splines \cite{friedman1991multivariate}. 5) Finally, we also consider a cross-fitted TMLE (CV-TMLE), which typically results in better finite sample performance due to avoidance of the Donkser class condition required for asymptotic normality \cite{klaassen1987consistent,zheng2011cross, chernozhukov2018double}.

\section{Results}
\begin{figure}[H]
    \centering
    \caption{Reliability function for the absolute bias of each of the four estimators (IPTW, g-computation, BART, TMLE, and CV-TMLE) when considering data generating mechanisms characterized by 1 binary treatment, 1 binary outcome, 5 binary covariates, 1 numeric covariate, and various levels of model complexity.}
    \label{fig:bias}
    \begin{subfigure}[]{\textwidth}
         \centering
         \includegraphics[width=.8\textwidth]{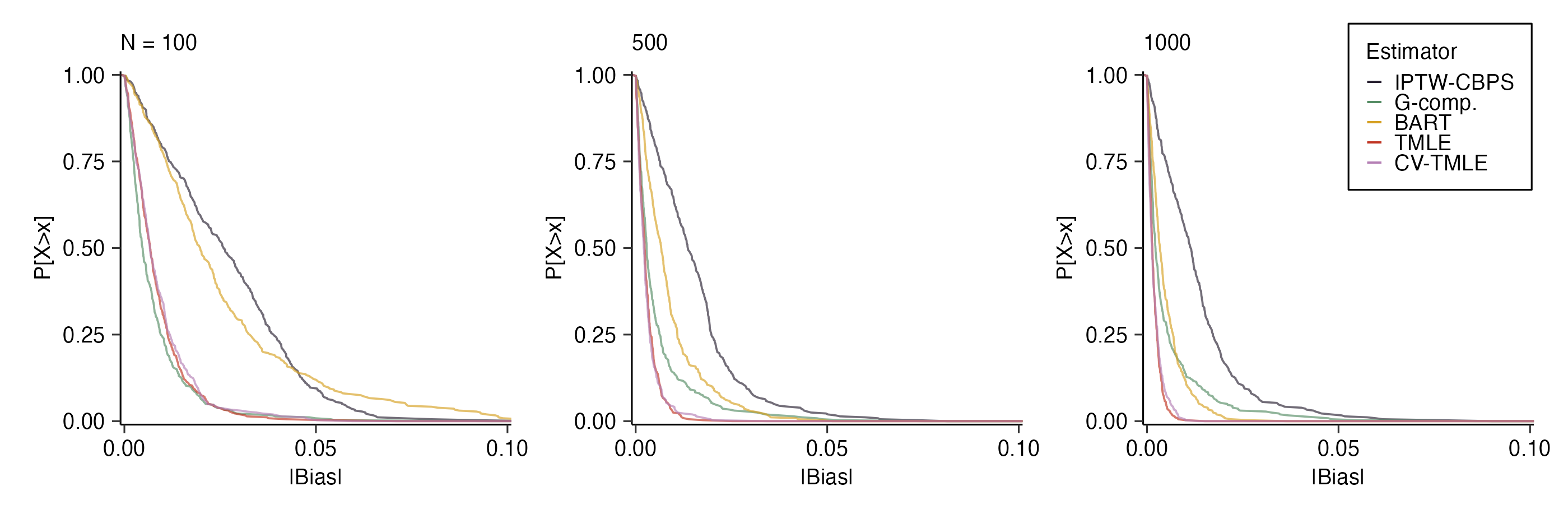}
         \caption{No treatment effect heterogeneity, no interactions}
         \label{fig:bias:a}
     \end{subfigure}
       \hfill
      \begin{subfigure}[]{\textwidth}
         \centering
         \includegraphics[width=.8\textwidth]{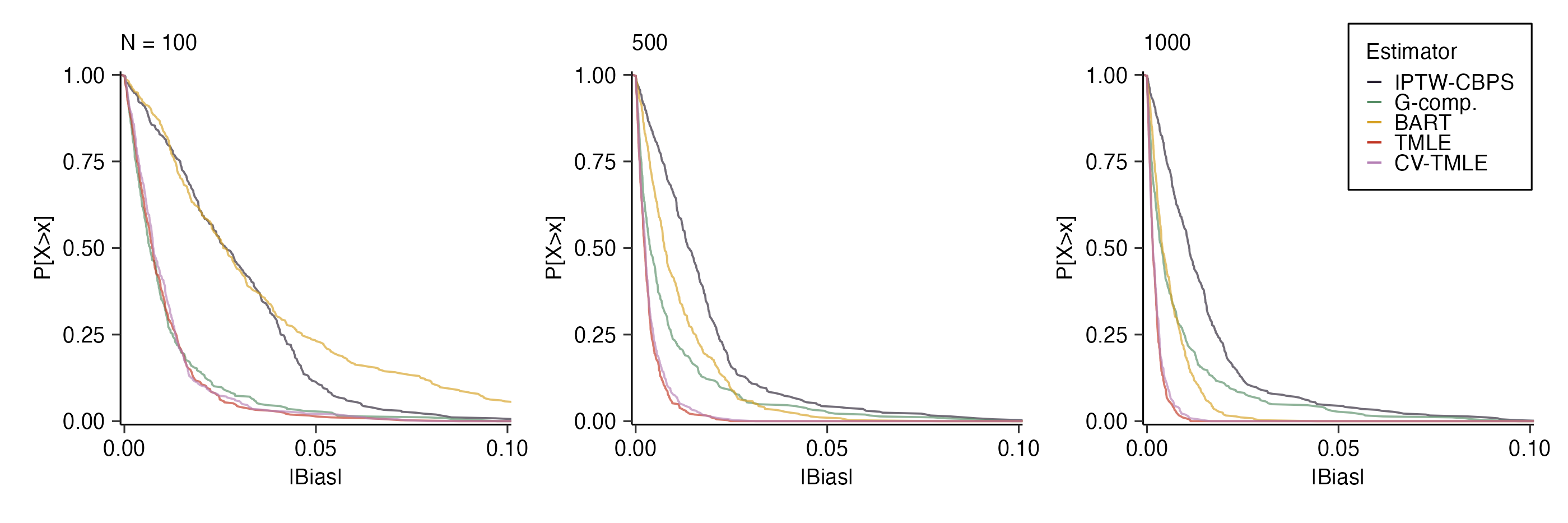}
         \caption{Treatment effect heterogeneity, no interactions}
         \label{fig:bias:b}
      \end{subfigure}
       \begin{subfigure}[]{\textwidth}
         \centering
         \includegraphics[width=.8\textwidth]{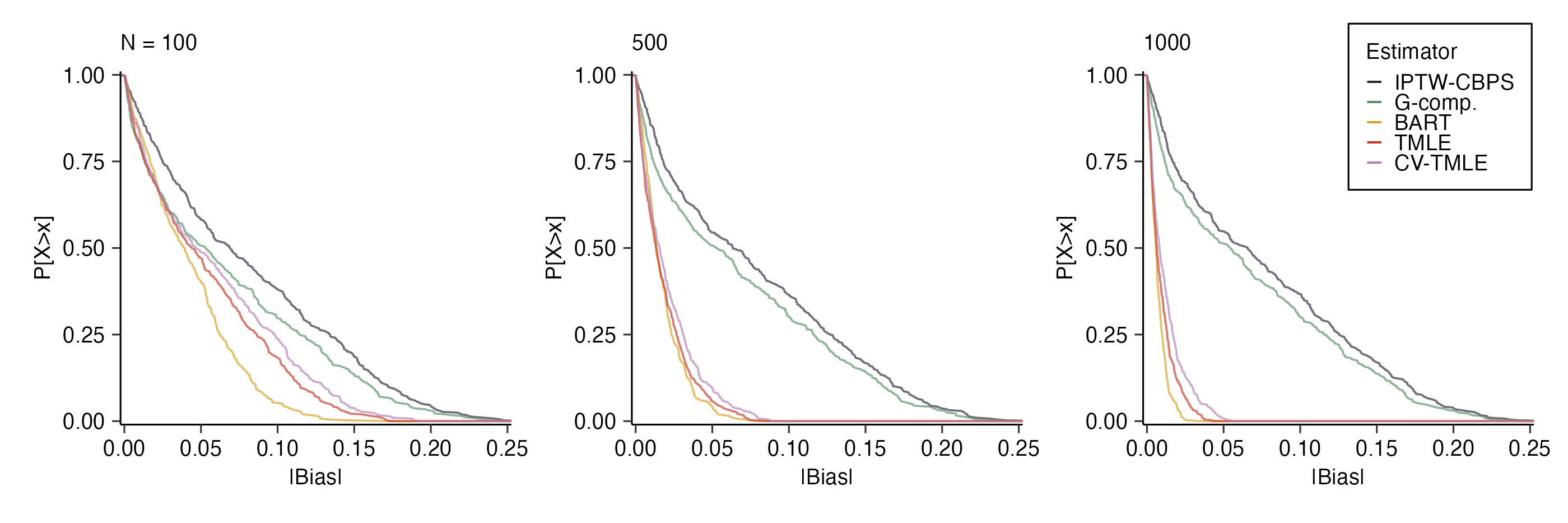}
         \caption{No treatment effect heterogeneity, up to 3-way interactions}
         \label{fig:bias:c}
     \end{subfigure}
       \hfill
      \begin{subfigure}[]{\textwidth}
         \centering
         \includegraphics[width=.8\textwidth]{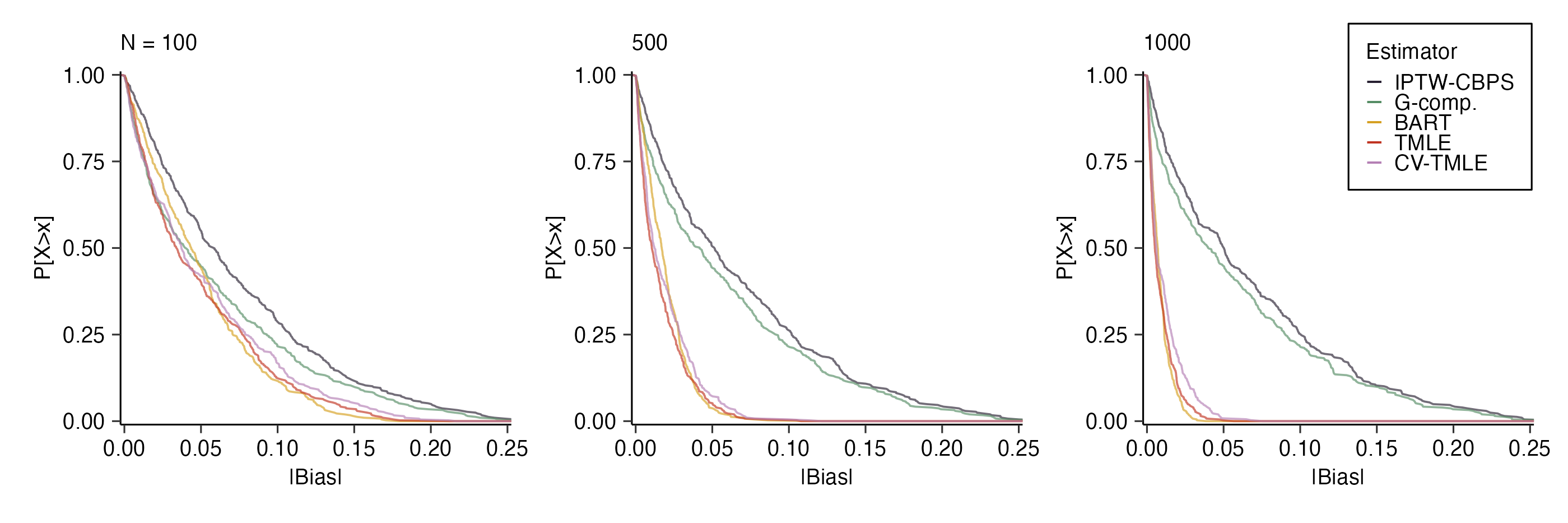}
         \caption{Treatment effect heterogeneity, up to 3-way interactions}
              \label{fig:bias:d}
      \end{subfigure}
\end{figure}

\begin{figure}[H]
    \centering
    \caption{Reliability function for the absolute MSE of each of the four estimators (IPTW, g-computation, BART,TMLE, and CV-TMLE) when considering data generating mechanisms characterized by 1 binary treatment, 1 binary outcome, 5 binary covariates, 1 numeric covariate, and various levels of model complexity.}
    \label{fig:mse}
   \begin{subfigure}[]{\textwidth}
         \centering
         \includegraphics[width=.8\textwidth]{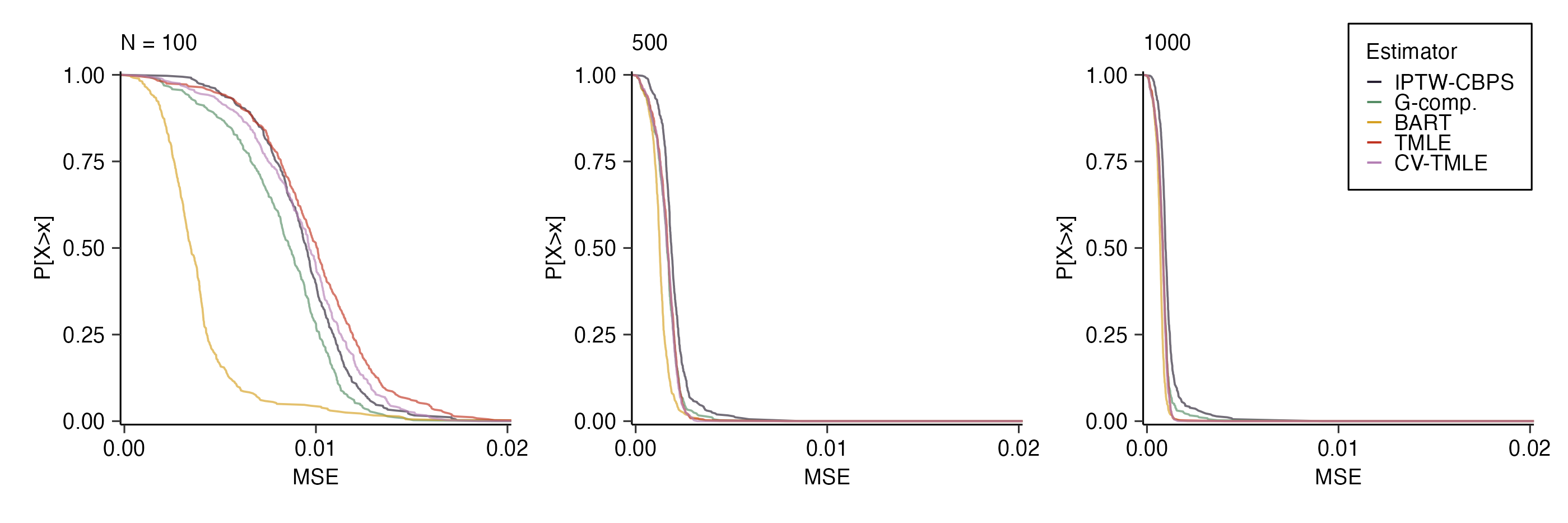}
         \caption{No treatment effect heterogeneity, no interactions}
         \label{fig:mse:a}
     \end{subfigure}
       \hfill
      \begin{subfigure}[]{\textwidth}
         \centering
         \includegraphics[width=.8\textwidth]{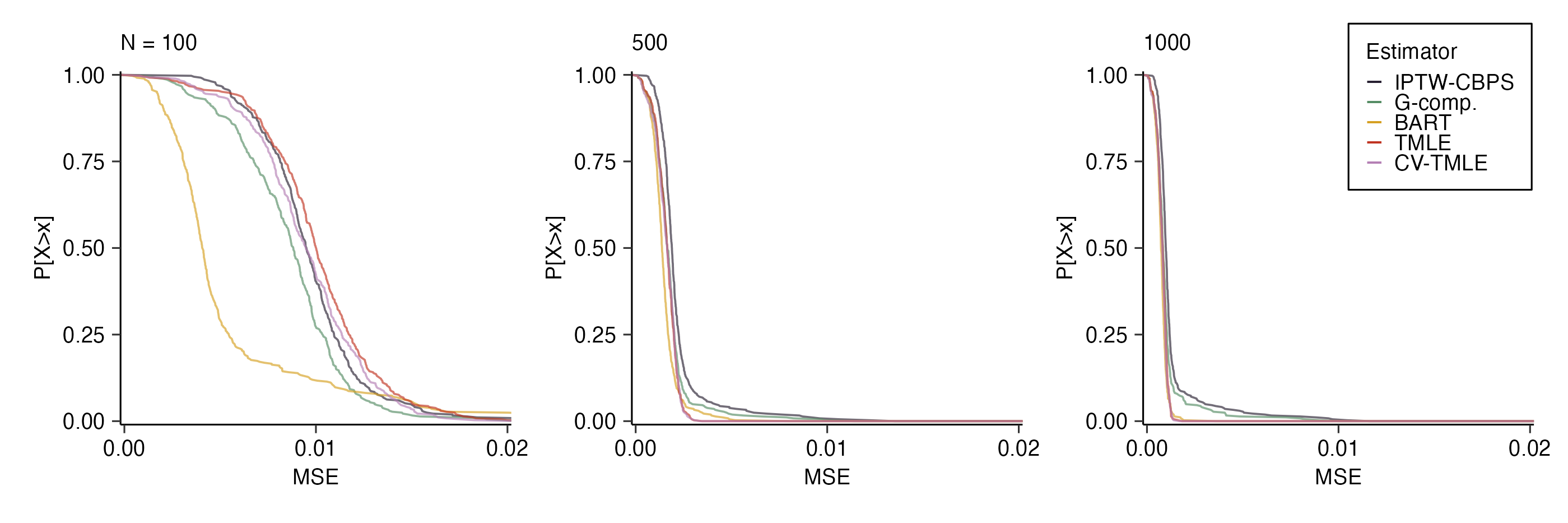}
         \caption{Treatment effect heterogeneity, no interactions}
         \label{fig:mse:b}
      \end{subfigure}
       \begin{subfigure}[]{\textwidth}
         \centering
         \includegraphics[width=.8\textwidth]{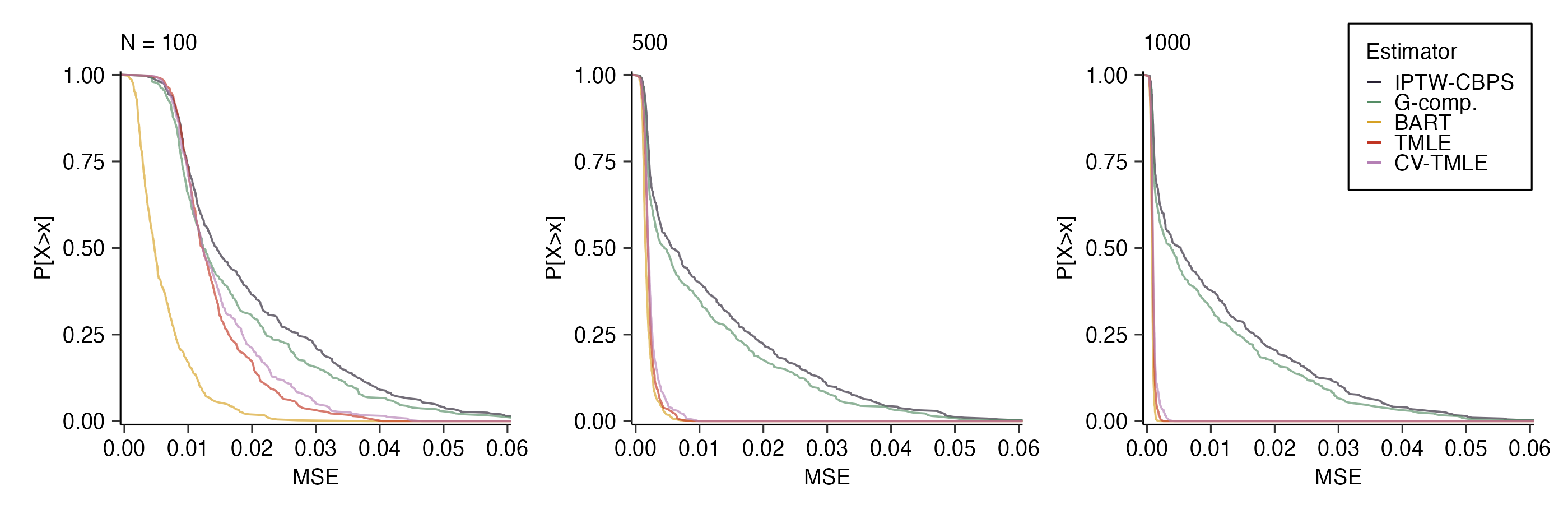}
         \caption{No treatment effect heterogeneity, up to 3-way interactions}
         \label{fig:mse:c}
     \end{subfigure}
       \hfill
      \begin{subfigure}[]{\textwidth}
         \centering
         \includegraphics[width=.8\textwidth]{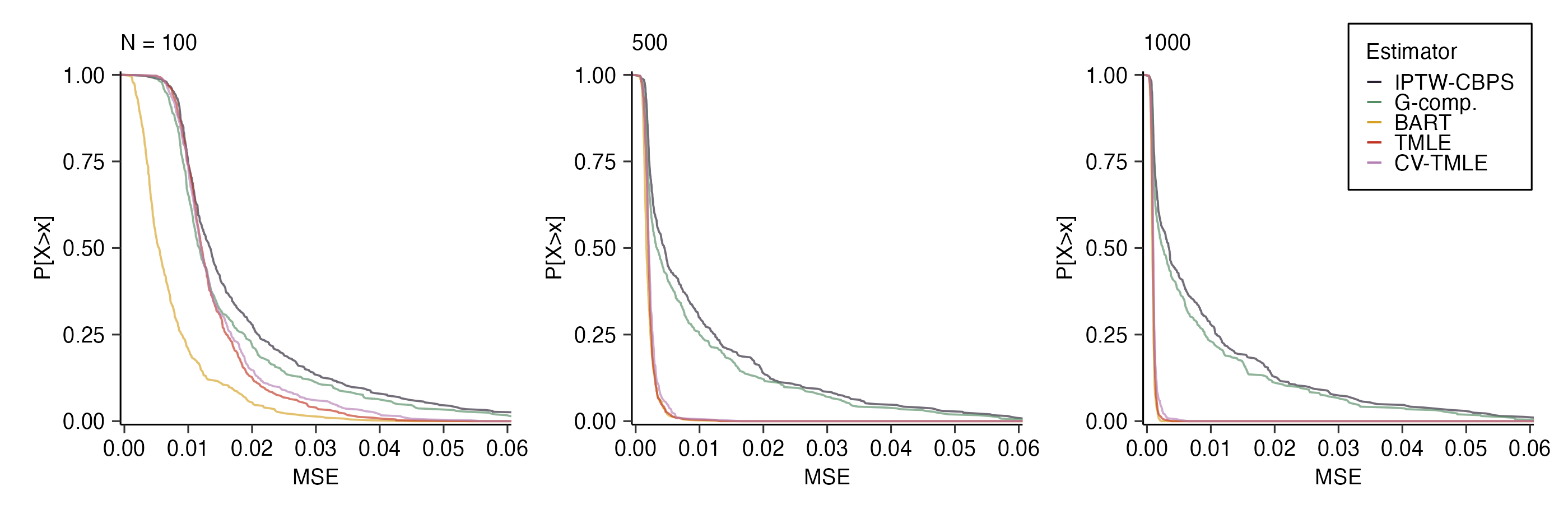}
         \caption{Treatment effect heterogeneity, up to 3-way interactions}
              \label{fig:mse:d}
      \end{subfigure}
\end{figure}

We evaluated estimator performance in terms of absolute bias, MSE, and 95\% CI coverage. Figures \ref{fig:bias} and \ref{fig:mse} show the reliability function for the proportion of $J$ data distributions where the estimator's absolute bias and MSE, respectively, are greater than $x$, for each value on the x-axis. Table \ref{tab:cov} gives the 95\% CI coverage. 
We stratify results by each estimator, the three sample sizes, and by complexity of the data-generating mechanism. To vary complexity we examine a) no interactions between variables and no treatment effect heterogeneity, b) treatment effect heterogeneity but no interactions between variables, c) up to 3rd order interactions between variables and no treatment effect heterogeneity, and d) up to 3rd order interactions and treatment effect heterogeneity. We also stratify results by degree of practical violations of the positivity assumption.  

We see in Figures \ref{fig:bias} and \ref{fig:mse} that at least one, and often, all three of the nonparametric estimators, BART, TMLE, and CV-TMLE perform better than both parametric estimators in terms of both bias and MSE. The one exception is for the smallest sample size of $N=100$ and the simplest two data-generating mechanisms that do not have any interactions between variables, the parametric g-computation estimator performs similar to the two TMLEs in terms of bias (Figures \ref{fig:bias:a} and \ref{fig:bias:b}), and BART performs relatively worse. In terms of MSE, however, at least one of the nonparametric estimators performs best across all settings (Figure \ref{fig:mse}), with BART dominating in all cases when $N=100$ and performing at least close to the best in all other scenarios. 

The largest separation between the parametric versus nonparametric reliability curves in Figures \ref{fig:bias} and \ref{fig:mse} occur for sample sizes $N \in \{500,1000\}$ and in the more complex settings of variable interactions and possibly treatment effect heterogeneity. For example, in the most complex setting with $N=1000$ (Figure \ref{fig:bias:d}), the nonparametric estimators result in absolute bias $>0.1$ in 0\% of data distributions, but the parametric estimators result in absolute bias $>0.1$ in 11.1\% and 13.3\% of data distributions for the parametric g-computation estimator and IPTW estimator, respectively. This result was anticipated, because in larger sample sizes, the asymptotic advantages of nonparametric estimators to model complexity and nonlinearities discussed in Section \ref{sec:theory} may take hold. 


We also stratify by degree of practical positivity violations (minimal, moderate, severe) in Figures S2 and S3 in the SI Appendix. Although estimator performance degrades slightly in the presence of severe positivity violations, particularly for the IPTW estimator, the relative performance of the estimators remains the same.

In terms of 95\% CI coverage (Table 1), we see that all estimators perform well for sample size $N=1000$ in the simplest scenarios without interactions, with slight over-coverage by IPTW and BART and slight under-coverage by the non-cross-fitted TMLE. However, with more complexity in terms of interactions, coverage suffers for the parametric estimators, particularly with increasing sample size; in contrast, BART and CV-TMLE continue to attain at least the nominal coverage rate. The recovery of nominal coverage by CV-TMLE compared to TMLE illustrates the importance of cross-fitting when using data-adaptive regression estimators for asymptotic normality \cite{klaassen1987consistent,zheng2011cross, chernozhukov2018double}. Interestingly, BART overcovers in every scenario despite generally having the smallest variance across scenarios. 


\begin{table*}[bt!]
\centering
 \caption{Median (IQR) values of the 95\% confidence interval coverage distributions across all estimators and sample sizes. Int = interaction, HTE = treatment effect heterogeneity.}
 \label{tab:cov}
\begin{tabular}[t]{cccccc}
\toprule
 N & IPTW & G comp. & BART & TMLE & CV-TMLE\\
\midrule
\multicolumn{5}{l}{\textbf{No int., no HTE}}\\
100 & 0.96 (0.95, 0.98) & 0.93 (0.92, 0.94) & 1.00 (0.99, 1.00) & 0.82 (0.80, 0.85)  & 0.97 (0.96, 0.98) \\
500 & 0.96 (0.95, 0.98) & 0.94 (0.94, 0.96) & 0.99 (0.98, 0.99) & 0.92 (0.90, 0.93)  & 0.96 (0.95, 0.96) \\
1000 & 0.96 (0.94, 0.98) & 0.95 (0.93, 0.96) & 0.98 (0.98, 0.99) & 0.93 (0.92, 0.94) & 0.95 (0.94, 0.96)\\
\midrule
\multicolumn{5}{l}{\textbf{No int., HTE}}\\
100 & 0.96 (0.95, 0.98) & 0.92 (0.91, 0.94) & 1.00 (0.99, 1.00) & 0.83 (0.80, 0.85)  & 0.97 (0.96, 0.98) \\
500 & 0.96 (0.95, 0.98) & 0.94 (0.93, 0.95) & 0.98 (0.97, 0.99) & 0.91 (0.90, 0.93)  & 0.96 (0.95, 0.96) \\
1000 & 0.96 (0.94, 0.98) & 0.94 (0.92, 0.95) & 0.98 (0.97, 0.99) & 0.93 (0.91, 0.94) & 0.96 (0.94, 0.96)\\
\midrule
\multicolumn{5}{l}{\textbf{2-way int., no HTE}}\\
100 & 0.93 (0.83, 0.96) & 0.91 (0.80, 0.93) & 0.99 (0.98, 1.00) & 0.78 (0.71, 0.81)  & 0.96 (0.93, 0.97) \\
500 & 0.86 (0.38, 0.95) & 0.87 (0.37, 0.94) & 0.98 (0.98, 0.99) & 0.89 (0.87, 0.91)  & 0.95 (0.94, 0.96) \\
1000 & 0.74 (0.10, 0.94) & 0.80 (0.11, 0.93) & 0.98 (0.98, 0.99) & 0.91 (0.90, 0.92) & 0.95 (0.94, 0.96)\\
\midrule
\multicolumn{5}{l}{\textbf{2-way int., HTE}}\\
100 & 0.94 (0.87, 0.96) & 0.91 (0.85, 0.94) & 0.99 (0.98, 1.00) & 0.79 (0.73, 0.83)  & 0.96 (0.94, 0.97) \\
500 & 0.88 (0.48, 0.95) & 0.89 (0.52, 0.94) & 0.98 (0.97, 0.99) & 0.89 (0.87, 0.90)  & 0.95 (0.94, 0.96) \\
1000 & 0.80 (0.17, 0.94) & 0.83 (0.22, 0.93) & 0.98 (0.97, 0.98) & 0.91 (0.90, 0.92) & 0.95 (0.94, 0.96)\\
\midrule
\multicolumn{5}{l}{\textbf{3-way int., no HTE}}\\
100 & 0.91 (0.77, 0.95) & 0.89 (0.76, 0.93) & 0.99 (0.97, 1.00) & 0.76 (0.65, 0.81)  & 0.96 (0.91, 0.97) \\
500 & 0.70 (0.20, 0.94) & 0.74 (0.22, 0.94) & 0.98 (0.96, 0.99) & 0.87 (0.80, 0.89)  & 0.94 (0.90, 0.96) \\
1000 & 0.47 (0.02, 0.93) & 0.55 (0.03, 0.93) & 0.98 (0.97, 0.99) & 0.89 (0.86, 0.91) & 0.94 (0.91, 0.96)\\
\midrule
\multicolumn{5}{l}{\textbf{3-way int., HTE}}\\
100 & 0.92 (0.84, 0.95) & 0.90 (0.82, 0.93) & 0.99 (0.96, 1.00) & 0.78 (0.70, 0.82)  & 0.96 (0.93, 0.97) \\
500 & 0.82 (0.38, 0.94) & 0.84 (0.42, 0.94) & 0.98 (0.96, 0.98) & 0.87 (0.82, 0.90)  & 0.94 (0.90, 0.96) \\
1000 & 0.66 (0.10, 0.93) & 0.70 (0.13, 0.93) & 0.98 (0.97, 0.98) & 0.90 (0.87, 0.92) & 0.94 (0.92, 0.96)\\
\bottomrule
\end{tabular}
\end{table*}


\section{Conclusions}
We proposed a Universal Monte-Carlo Simulation method to bring evidence to the debate \cite{imbens2004nonparametric} as to whether nonparametric estimators that use data-adaptive machine learning algorithms in model fitting confer any meaningful advantage over simpler parametric methods in real-world finite sample analyses. Previously, others sought to contribute finite sample evidence in favor of one estimator or class of estimators over another by conducting simulation studies \cite{dorie2019automated,porter2011relative,ozery2018adversarial,parikh2022validating,schuler2017synth}. However, simulation studies are limited in that they only evaluate estimator performance across a small number of data-generating mechanisms, which may not be representative of performance in general. 
Consequently, our proposed approach greatly expands the number of data generating mechanisms considered from a small few to thousands, which is likely to result in more generalizable---and thus, informative---evidence of estimator performance.

We applied our proposed approach to compare performance of nonparametric estimators, BART, TMLE, and CV-TMLE, to two parametric estimators, outcome regression-based g-computation and IPTW, in finite samples of sizes N=100, 500, and 1000 and across different degrees of model complexity. In doing so, we provide what is to our knowledge the first general evidence of nonparametric versus parametric performance across a universe of possible research settings and in finite samples where asymptotic properties learned from theoretical results may not provide good approximations. We found that even in small samples, nonparametric estimators nearly always outperform the parametric estimators in terms of bias and MSE. However, the advantage of nonparametric estimation attenuated with decreasing sample size and decreasing complexity, and in the simplest data-generating mechanisms and samples of $N=100$, the parametric g-computation estimator performed similarly to nonparametric TML estimator in terms of bias and between BART and TMLE in terms of MSE. 

Even though our results were learned from 2,250,000 data sets across 3,000 data generating mechanisms, the space of data-generating mechanisms we considered was nonetheless limited. In particular, we only considered settings with a binary treatment, a binary outcome, and six confounding variables, only one of which was multi-valued. It is certainly possible that our conclusions would differ for more complex settings. In future work, we will develop a software tool for running simulations with user-specified outcome and covariate types and covariate dimensions. This way, users would be able to compare estimator performances over a space of data generating mechanisms that are likely to contain the probability distribution corresponding to their real-world data sampling setting or a close approximation thereof.

Lastly, although we have shown that the choice of nonparametric vs. parametric estimator may matter---in some cases more than others---all estimators are limited by the data input. Unmeasured variables and variables measured with error are significant and near-ubiquitous limitations that can thwart accurate causal effect estimation and inference. 
A relatively recent high-profile and high-stakes example involved data errors leading to inaccurate algorithmic predictions in the criminal justice system that resulted in unintended parole denials \cite{rudin2019secrets,wexler2017computer}. We can work to improve the estimation step of answering research questions, but the accuracy of our answers will be limited by the weakest link, highlighting the importance of theory, subject matter knowledge, identification, and data quality, in addition to estimation. 

In this article, we have focused on strengthening the estimation link. Our results show that in the large space of settings we have considered, this can be accomplished by employing nonparametric estimators grounded in asymptotic theory to substantially reduce bias in large-sample settings with interactions and nonlinearities while compromising very little in terms of performance even in simple, small-sample settings.
\newpage

\begin{appendix}

\large{\textbf{Supplementary Materials for \titlepaper}}
\vspace{1cm}
\renewcommand\thefigure{S\arabic{figure}}
\renewcommand{\tablename}{Table S}
\renewcommand{\thesection}{S\arabic{section}}
\renewcommand{\thesubsection}{S\arabic{subsection}}
\setcounter{table}{0} 
\setcounter{figure}{0}  
\subsection*{Checking confounding bias}
The confounding bias for a given distribution is equal to
\begin{equation}
\begin{aligned}
     C &= \sum_t(2t-1)\sum_x\frac{\P(X=x)}{\P(T=t)} \left\{\P(T=t\mid X=x) -
    \P(T=t)\right\}\\
    &\P(Y=1\mid T=t, X=x),\label{eq:conf}
\end{aligned}
\end{equation}
so we must ensure that $\P(X=x)$ and $\P(T=t\mid X=x)$ are such
that it is possible to find values $0\leq\P(Y=1\mid T=t, X=x)\leq 1$
such that $C=b$. This occurs if
$C_{\text{low}}\leq b \leq C_{\text{high}}$, where
\begin{align*}
  C_{\text{high}} &= \sum_x\max\left\{\frac{\P(X=x)}{\P(T=1)}\left\{\P(T=1\mid X=x) -
                   \P(T=1)\right\}, 0\right\}\\
  &-\sum_x\min\left\{\frac{\P(X=x)}{\P(T=0)}\left\{\P(T=0\mid X=x) -
  \P(T=0)\right\}, 0\right\}\\
  C_{\text{low}} &= \sum_x\min\left\{\frac{\P(X=x)}{\P(T=1)}\left\{\P(T=1\mid X=x) -
                    \P(T=1)\right\}, 0\right\}\\
                  &-\sum_x\max\left\{\frac{\P(X=x)}{\P(T=0)}\left\{\P(T=0\mid X=x) -
                    \P(T=0)\right\}, 0\right\}.
\end{align*}
If the condition $C_{\text{low}}\leq b \leq C_{\text{high}}$ is false,
we reject the current draw of $\{\P(X=x):x\in \mathcal X\}$, and
$\{(f_l, \alpha_{0,l}, \alpha_{1,l}):l\in L\}$ and repeat the process until
$C_{\text{low}}\leq b \leq C_{\text{high}}$. If this condition is not
achievable after 1000 iterations, this is an indicator that the
initial conditions may be infeasible. In this case the algorithm fails
and does not return a sampled distribution. 

\begin{figure}
    \centering
    \caption{Relationship between probability of treatment conditional on the numerical covariate and the values of the numerical covariate for different values of $\eta$ and $\rho$.}
    \label{fig:nonlinear}
\includegraphics[width=.8\textwidth]{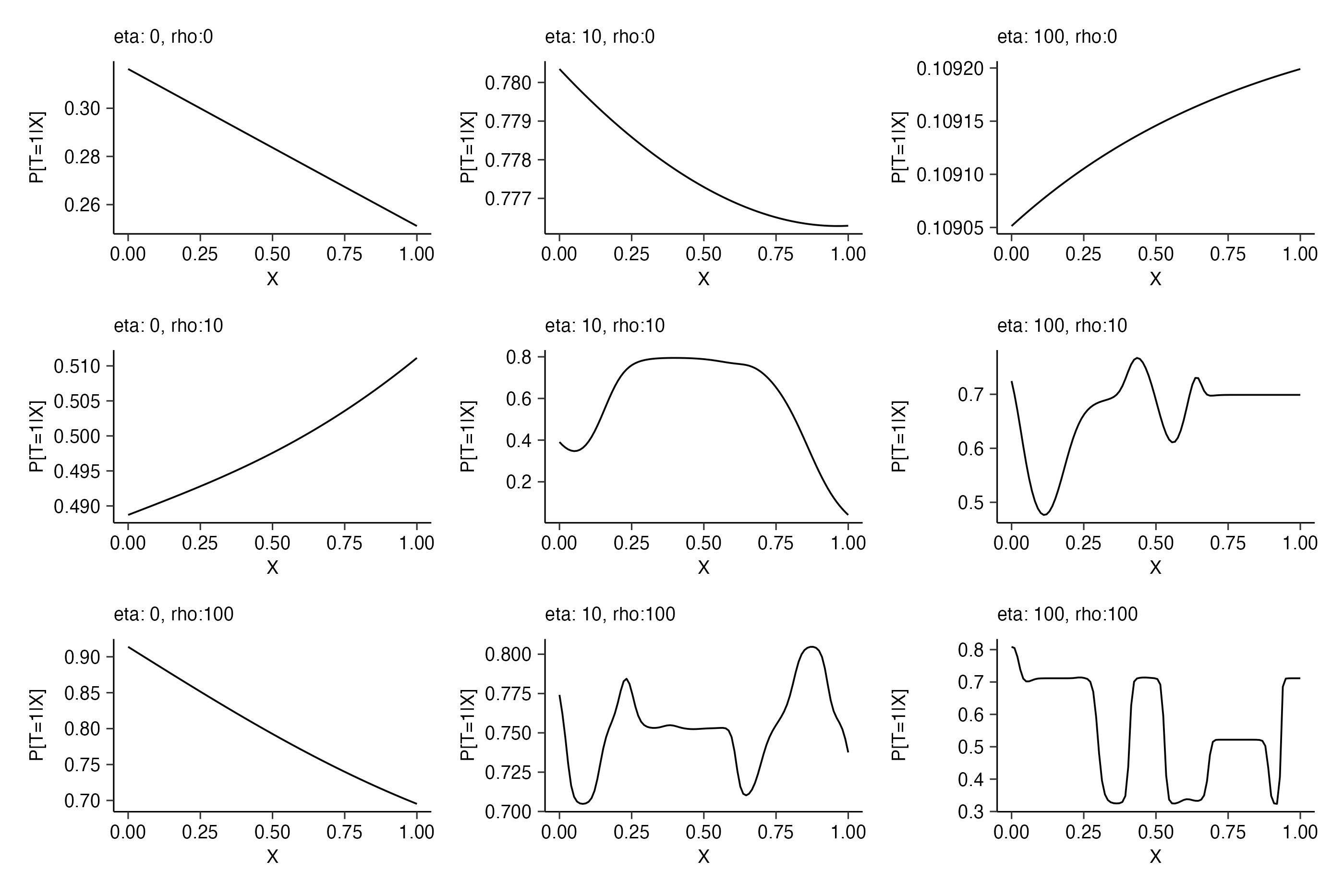}

\end{figure}

\begin{figure}
    \centering
    \caption{Reliability function for the absolute bias of each of the four estimators (CBPS, g-computation, BART, and TMLE) by degree of positivity violations when considering data generating mechanisms characterized by 1 binary treatment, 1 binary outcome, 5 binary covariates, 1 numeric covariate, and various levels of model complexity.}
    \label{fig:bias:pos}
    \begin{subfigure}[]{\textwidth}
         \centering
         \includegraphics[width=.8\textwidth]{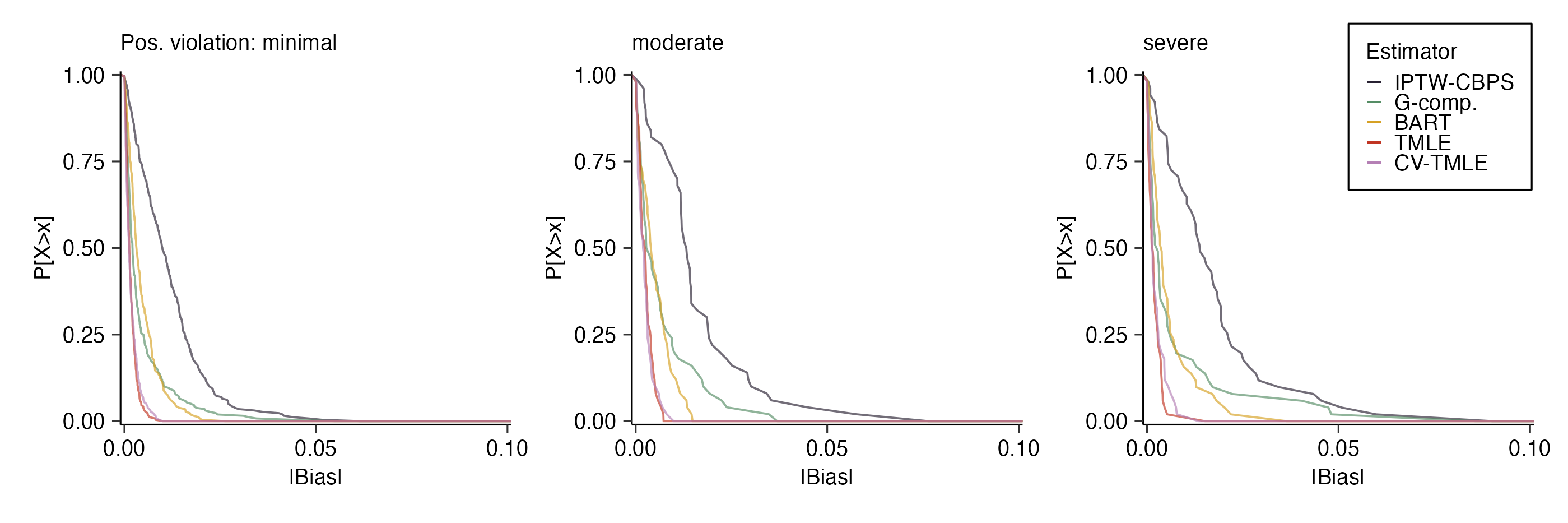}
         \caption{No treatment effect heterogeneity, no interactions}
         \label{fig:bias:pos:a}
     \end{subfigure}
       \hfill
      \begin{subfigure}[]{\textwidth}
         \centering
         \includegraphics[width=.8\textwidth]{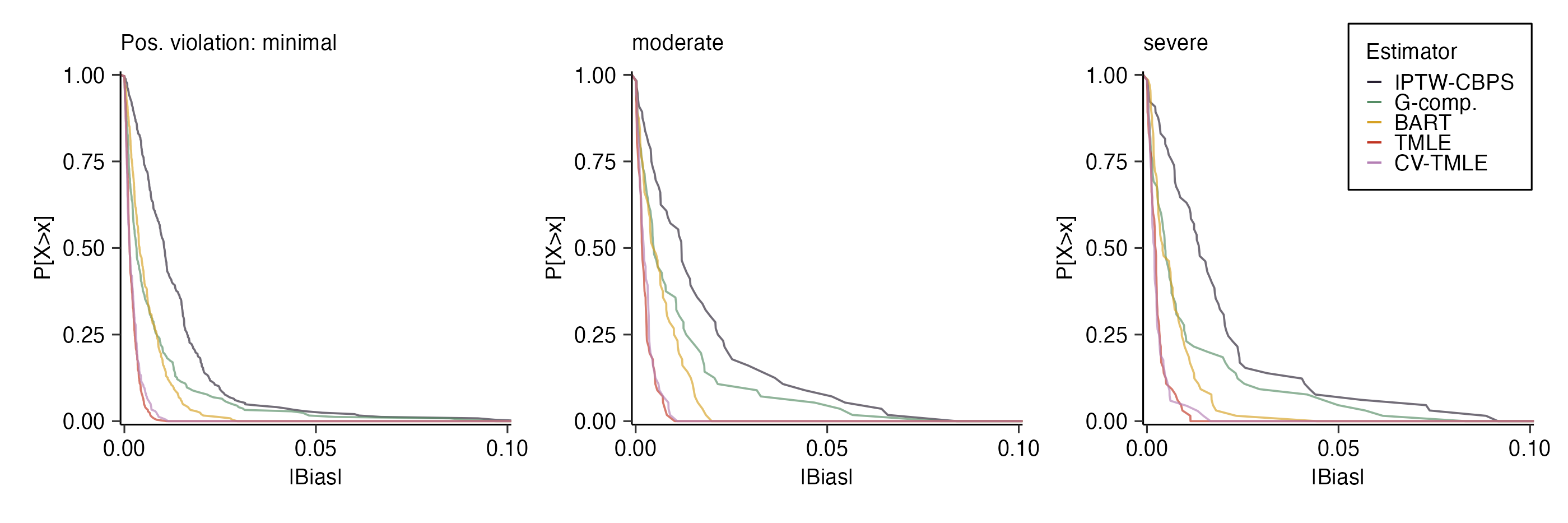}
         \caption{Treatment effect heterogeneity, no interactions}
         \label{fig:bias:pos:b}
      \end{subfigure}
       \begin{subfigure}[]{\textwidth}
         \centering
         \includegraphics[width=.8\textwidth]{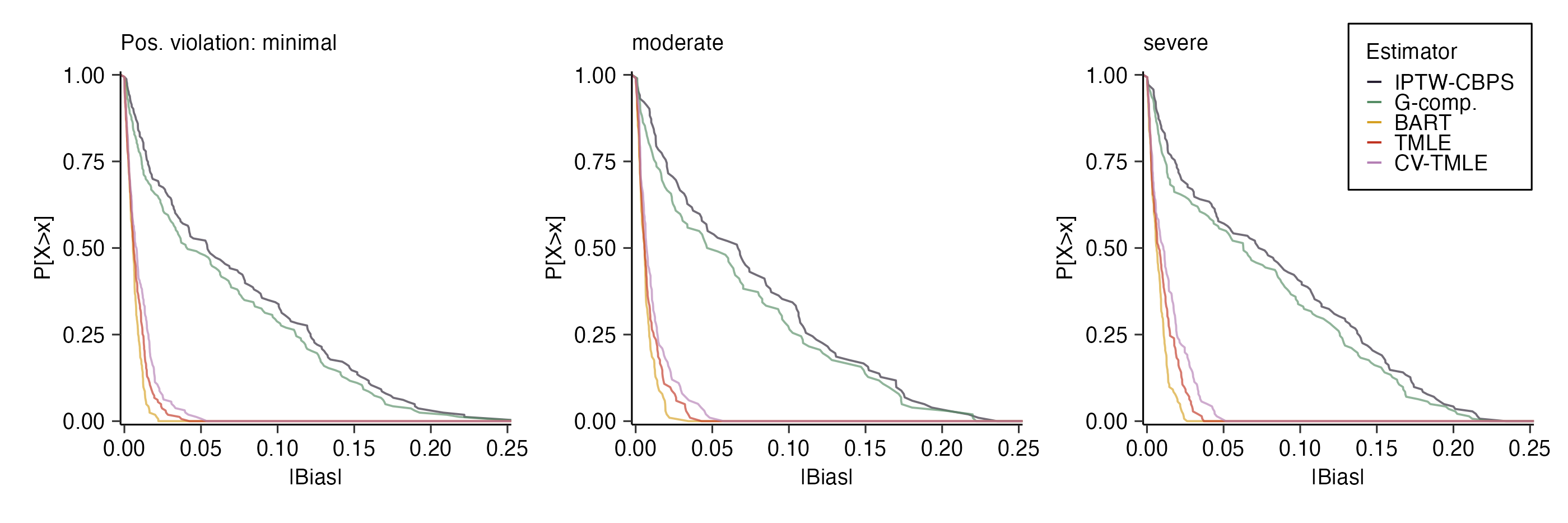}
         \caption{No treatment effect heterogeneity, up to 3-way interactions}
         \label{fig:bias:pos:c}
     \end{subfigure}
       \hfill
      \begin{subfigure}[]{\textwidth}
         \centering
         \includegraphics[width=.8\textwidth]{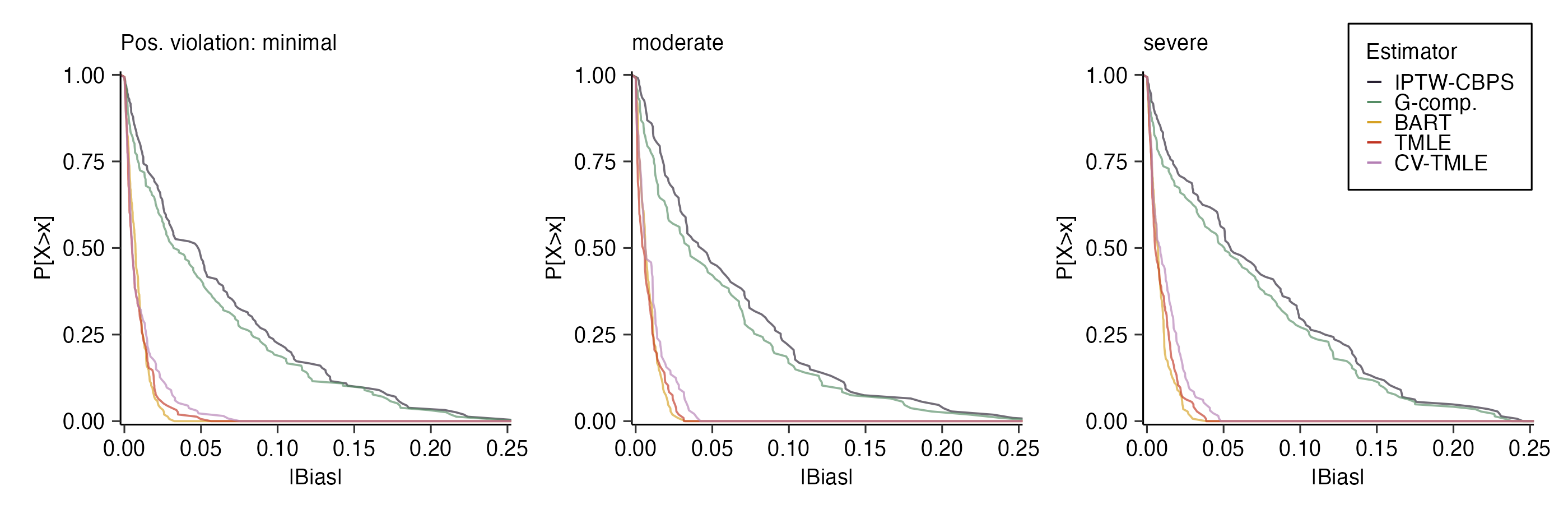}
         \caption{Treatment effect heterogeneity, up to 3-way interactions}
              \label{fig:bias:pos:d}
      \end{subfigure}
\end{figure}

\begin{figure}
    \centering
    \caption{Reliability function for the absolute MSE of each of the four estimators (IPTW, g-computation, BART, and TMLE) by degree of positivity violations when considering data generating mechanisms characterized by 1 binary treatment, 1 binary outcome, 5 binary covariates, 1 numeric covariate, and various levels of model complexity.}
    \label{fig:mse:pos}
   \begin{subfigure}[]{\textwidth}
         \centering
         \includegraphics[width=.8\textwidth]{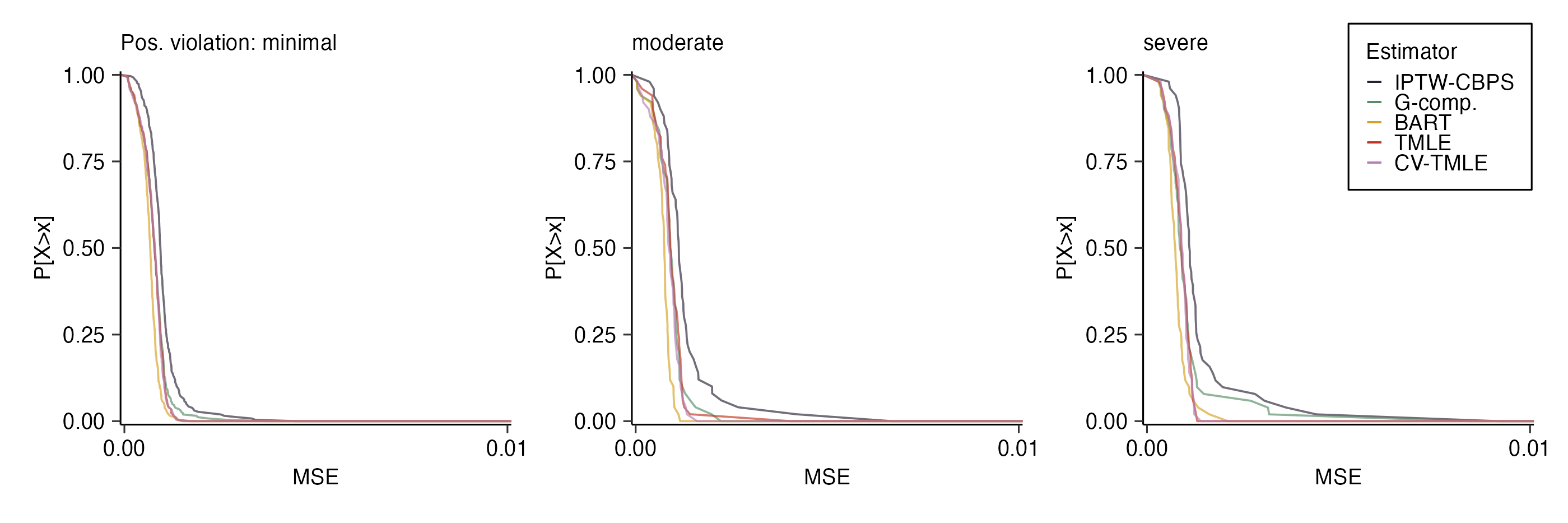}
         \caption{No treatment effect heterogeneity, no interactions}
         \label{fig:mse:pos:a}
     \end{subfigure}
       \hfill
      \begin{subfigure}[]{\textwidth}
         \centering
         \includegraphics[width=.8\textwidth]{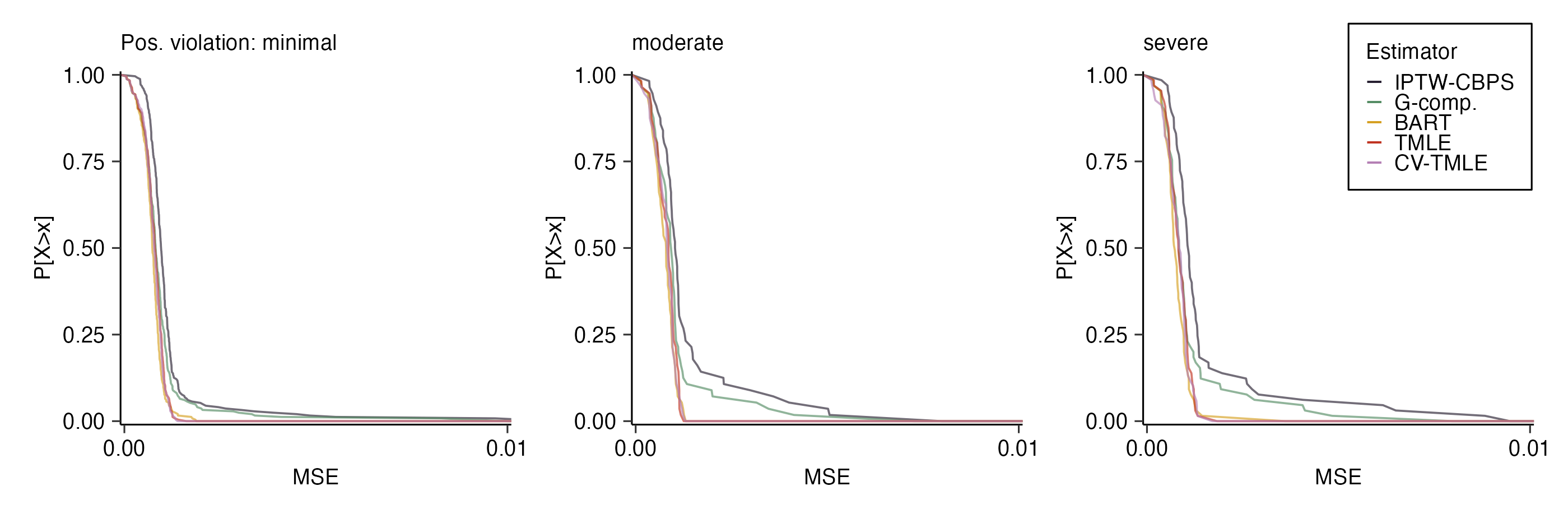}
         \caption{Treatment effect heterogeneity, no interactions}
         \label{fig:mse:pos:b}
      \end{subfigure}
       \begin{subfigure}[]{\textwidth}
         \centering
         \includegraphics[width=.8\textwidth]{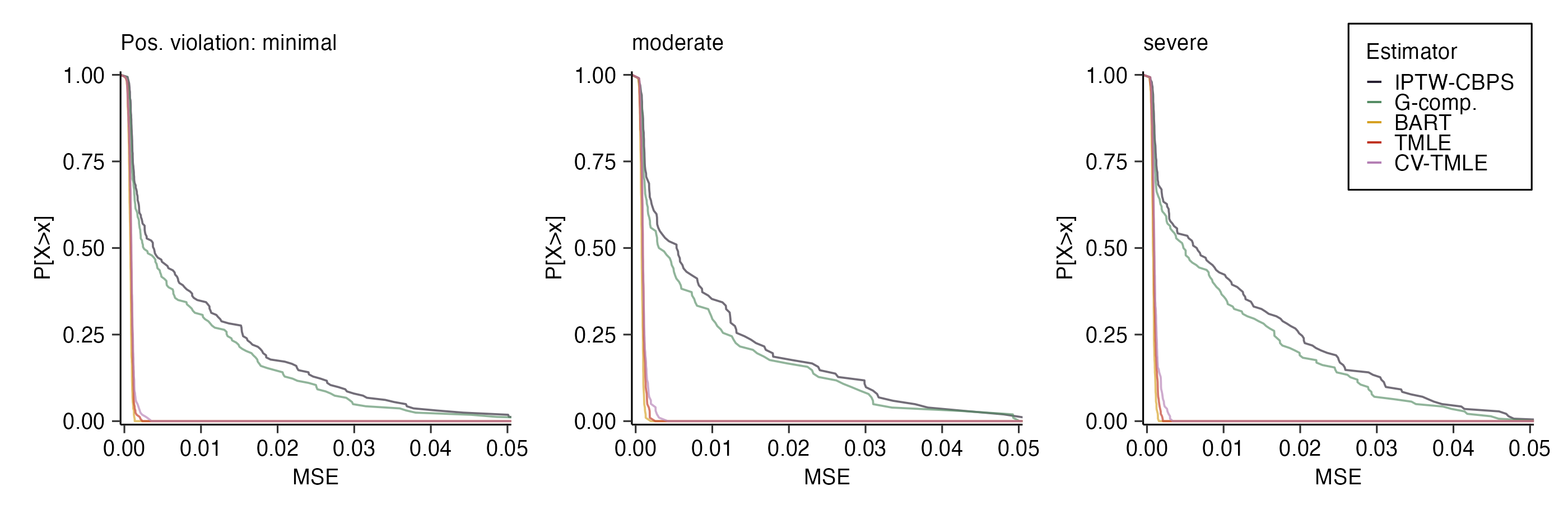}
         \caption{No treatment effect heterogeneity, up to 3-way interactions}
         \label{fig:mse:pos:c}
     \end{subfigure}
       \hfill
      \begin{subfigure}[]{\textwidth}
         \centering
         \includegraphics[width=.8\textwidth]{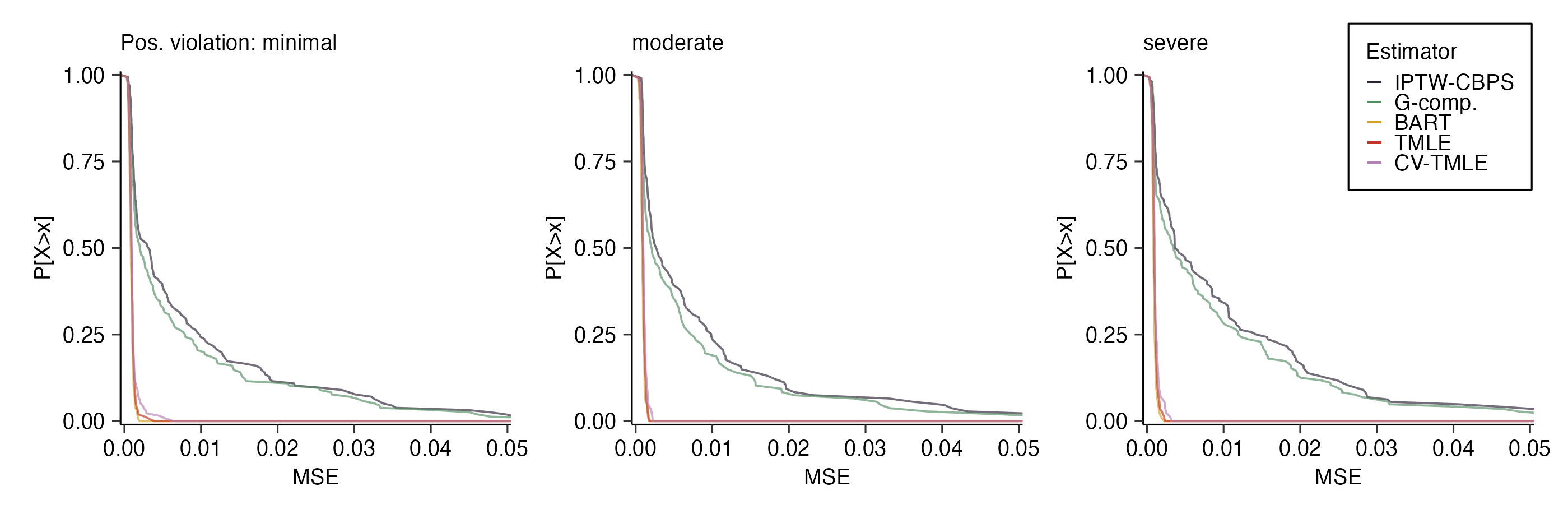}
         \caption{Treatment effect heterogeneity, up to 3-way interactions}
              \label{fig:mse:pos:d}
      \end{subfigure}
\end{figure}

\end{appendix}

\bibliographystyle{plainnat}
\bibliography{refs}
\end{document}

%% file: _preamble.tex
\usepackage[inline,shortlabels]{enumitem}
\usepackage{float}
\usepackage{mathtools}
\usepackage{graphicx}
\usepackage[round]{natbib}
\usepackage{bbm}
\usepackage{tikz}
\usepackage[english]{babel}
\usepackage{longtable}
\usepackage{color}
\usepackage{amssymb,amsmath,amsthm}
\usepackage{multirow}
\usepackage[titletoc,title]{appendix}
\usepackage{authblk}
\usepackage{setspace}
\usepackage{dsfont}
\usepackage[OT1]{fontenc}
\usepackage{refcount}
\graphicspath{ {./figs/} }

\usepackage{nameref,zref-xr} 
\zxrsetup{toltxlabel} 

\AtEndDocument{\refstepcounter{theorem}\label{finalthm}}
\AtEndDocument{\refstepcounter{equation}\label{finaleq}}

{
  \theoremstyle{definition}
  
}
{
  \theoremstyle{definition}
  
}
{
  \theoremstyle{definition}
  
}

\AtEndDocument{\refstepcounter{lemma}\label{finallemma}}

\DeclareMathOperator{\expit}{expit}

\DeclareMathOperator{\logit}{logit}

\renewcommand{\P}{\mathsf{P}}
\newcommand{\ate}{\mathsf{ATE}}

\newcommand{\mor}{\mathsf{OR}}

\newcommand{\mrr}{\mathsf{RR}}

\newcommand{\m}{\mathsf{m}}

\newcommand{\Bsub}{\mathsf{B}_{\mbox{\scriptsize sub}}}
\newcommand{\Btmle}{\mathsf{B}_{\mbox{\scriptsize tmle}}}

\newcommand{\Bn}{\mathsf{B}_l^{(n)}}
\newcommand{\MSEn}{\mathsf{MSE}_l^{(n)}}
\newcommand{\Covn}{\mathsf{CICov}_l^{(n)}}

\newcommand{\g}{\mathsf{g}}

\newcommand{\indep}{\mbox{$\perp\!\!\!\perp$}} 
 \newcommand{\dd}{\mathrm{d}}

\newcommand{\thetasub}{\hat\theta_{\mbox{\scriptsize sub}}(\delta)}

\newcommand{\thetatmle}{\hat\theta_{\mbox{\scriptsize tmle}}}
\newcommand{\thetaos}{\hat\theta_{\mbox{\scriptsize os}}}
 
\newcommand{\one}{\mathds{1}}
 
\newcommand{\E}{\mathsf{E}}

\newcommand{\sigmasub}{\sigma_{\mbox{\scriptsize sub}}}

\newcommand{\Xb}{X_{\text{bin}}}

\newcommand{\xbj}{x_{\text{bin}, j}}
\newcommand{\Xn}{X_{\text{num}}}
\newcommand{\xn}{x_{\text{num}}}
\newcommand{\xni}{x_{\text{num}, i}}
\newcommand{\xnj}{x_{\text{num}, j}}

\DeclarePairedDelimiterX{\norm}[1]{\lVert}{\rVert}{#1}
\usepackage{subcaption}
\usepackage{booktabs}
\pgfdeclarelayer{background}
\pgfsetlayers{background,main}
\usetikzlibrary{arrows,positioning}
\tikzset{
>=stealth',
punkt/.style={
rectangle,
rounded corners,
draw=black, very thick,
text width=6.5em,
minimum height=2em,
text centered},
pil/.style={
->,
thick,
shorten <=2pt,
shorten >=2pt,}
}
\newcommand{\Vertex}[2]
{\node[minimum width=0.6cm,inner sep=0.05cm] (#2) at (#1) {$\footnotesize#2$};
}
\newcommand{\Vertexr}[2]
{\node[rectangle, draw, minimum width=0.6cm,inner sep=0.05cm] (#2) at (#1) {$\footnotesize#2$};
}
\newcommand{\ArrowR}[3]%
{ \begin{pgfonlayer}{background}
\draw[->,#3] (#1) to[bend right=30] (#2);
\end{pgfonlayer}
}
\newcommand{\ArrowL}[3]%
{ \begin{pgfonlayer}{background}
\draw[->,#3] (#1) to[bend left=45] (#2);
\end{pgfonlayer}
}
\newcommand{\EdgeL}[3]%
{ \begin{pgfonlayer}{background}
\draw[dashed,#3] (#1) to[bend right=-45] (#2);
\end{pgfonlayer}
}

\newcommand{\Arrow}[3]%
{ \begin{pgfonlayer}{background}
\draw[->,#3] (#1) -- +(#2);
\end{pgfonlayer}
}
\RequirePackage[margin=1.25in,lines=32]{geometry}

\newcommand{\titlepaper}{All models are wrong, but which are useful? Comparing parametric and nonparametric estimation of causal effects in finite samples}
\date{}
\author[1]{Kara E.~Rudolph}
\author[1]{Nicholas Williams} 
\author[2]{Caleb H.~Miles}
\author[3]{Joseph Antonelli}
\author[4]{Iv\'an D\'iaz}

\affil[1]{\small Department of
  Epidemiology, Mailman School of Public Health, Columbia University.}
\affil[2]{\small Department of
  Biostatistics, Mailman School of Public Health, Columbia University.}
  \affil[2]{\small Department of
  Statistics, University of Florida.}
\affil[4]{\small Division of Biostatistics, Department of Population
  Health, New York University Grossman School of Medicine.}